\documentclass{phbauth}
\usepackage{graphicx}

\begin{document}

\begin{frontmatter}

\title{Phase diagram of the $\frac{1}{2}$-$\frac{1}{2}$-1-1 spin 
chain by the nonlinear $\sigma$ model}

\author{Ken'ichi Takano\thanksref{thank1}}

\address{
Laboratory of Theoretical Condensed Matter Physics and 
Research Center for Advanced Photon Technology, \\
Toyota Technological Institute, Nagoya 468-8511, Japan}

\thanks[thank1]{E-mail: takano@toyota-ti.ac.jp}

\begin{abstract}
   We examine a periodic mixed spin chain with spin magnitudes 
$\frac{1}{2}$ and 1 which are arrayed as
$\frac{1}{2}$-$\frac{1}{2}$-1-1. 
   The three independent parameters are ratios of the four exchange 
couplings. 
   We determine phase boundaries in the parameter space by using 
the gapless condition which was previously derived by mapping 
a general inhomogeneous spin chain to the nonlinear $\sigma$ 
model. 
   We find two gapless boundaries separating three disordered 
phases. 
   The features of the phases are explained in terms of singlet 
clusters. 
\end{abstract}

\end{frontmatter}

      Since the prediction of the Haldane gap for spin chains 
\cite{Haldane}, the nonlinear $\sigma$ model (NLSM) is considered 
to be a useful tool to investigate spin systems. 
      After then an NLSM for a spin chain with bond alternation 
is derived and examined \cite{Affleck1}. 
      The application of the NLSM to spin chains with more than one 
spin species is an interesting problem. 
      In a previous paper \cite{Takano1} we derived an NLSM for a 
general mixed quantum spin chain with arbitrary finite period. 
      However its various applications have not been examined 
in detail \cite{Fukui}. 

      In this paper we examine the case that the spin magnitudes 
in a unit cell are arrayed as $\frac{1}{2}$-$\frac{1}{2}$-1-1 
with spacing $a$. 
      The Hamiltonian is written as 
\begin{eqnarray}
      H &=& \sum_{j} 
 ( J \, {\bf S}_{4j+1} \cdot {\bf S}_{4j+2} 
 + J_{+} \, {\bf S}_{4j+2} \cdot {\bf S}_{4j+3} 
\nonumber \\
 &+&  {\bf S}_{4j+3} \cdot {\bf S}_{4j+4} 
 + J_{-} \, {\bf S}_{4j+4} \cdot {\bf S}_{4j+5} ) , 
\label{Hamiltonian}
\end{eqnarray}
where the magnitudes of ${\bf S}_{4j+1}$ and ${\bf S}_{4j+2}$ are 
$\frac{1}{2}$ and those of ${\bf S}_{4j+3}$ and ${\bf S}_{4j+4}$ 
are 1. 
      We reparametrize the coupling constants as 
\begin{eqnarray}
J_{+} = J'/(1+\delta) , \quad J_{-} = J'/(1-\delta) . 
\end{eqnarray}
      Following Ref. \cite{Takano1}, we map the spin chain described 
by Eq. (\ref{Hamiltonian}) to an NLSM. 
      The resultant NLSM action is given by 
\begin{eqnarray}
\label{action-NLSM}
      &{}& S_{\rm eff} = \int d\tau \int dx 
\biggl\{ 
- i \frac{J^{(0)}}{J^{(1)}} 
{\bf n} \cdot (\partial_{\tau} {\bf n} \times \partial_x {\bf n}) 
\nonumber \\ 
      &{}& + \frac{1}{2aJ^{(1)}} \biggl( 
\frac{J^{(1)}}{J^{(2)}} - \frac{J^{(0)}}{J^{(1)}} \biggl) 
(\partial_{\tau} {\bf n})^2 
+ \frac{a}{2} J^{(0)} (\partial_x {\bf n})^2 
\biggl\} , 
\nonumber \\ 
&{}& \frac{4}{J^{(0)}} = \frac{4}{J} + \frac{4}{J'} + 1 , 
\frac{4}{J^{(1)}} = \frac{2}{J} + 1 , 
\frac{4}{J^{(2)}} = \frac{1}{J} + 1 . 
\nonumber 
\end{eqnarray}

\begin{figure}[btp]
\begin{center}\leavevmode
\includegraphics[width=0.8\linewidth]{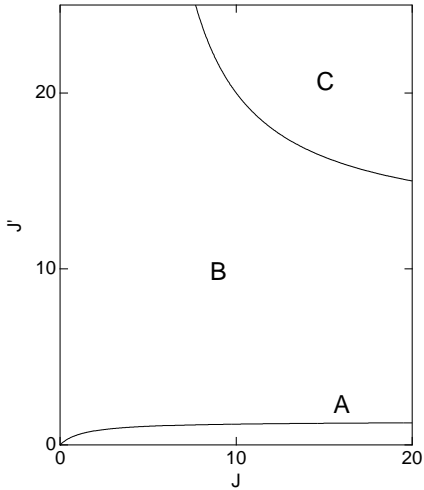}
\caption{ The phase diagram in the $J$-$J'$ plane 
for arbitrarily fixed $\delta$. 
Three disordered phases are separated by two gapless lines. }
\label{Fig_Phase}
\end{center}
\end{figure}

      The NLSM has gapless excitations if the 
topological angle $4\pi J^{(0)}/J^{(1)}$ is a half-odd-integer 
multiple of $2\pi$. 
      This condition is written as 
\begin{eqnarray}
\frac{1}{J'} = - \frac{2l-3}{2l-1} \frac{1}{J} 
- \frac{1}{4} \frac{2l-5}{2l-1} \quad (l=1, 2) . 
\label{gapless_line}
\end{eqnarray}
      This equation for each $l$ determines a phase boundary 
between gapful disordered phases. 
      Since the equation is independent of $\delta$, 
the phase boundaries in the $J$-$J'$-$\delta$ space is uniform 
in the $\delta$ direction. 
      We present the phase diagram in a $J$-$J'$ plane 
perpendicular to the $\delta$ axis in Fig.~\ref{Fig_Phase}. 
      In the special case of $J_{+}$ = $J_{-}$ (=$J'$), Tonegawa 
et al. performed a quantum Monte Carlo simulation and obtained a 
phase diagram \cite{Tonegawa1}.  
      For example, their phase boundary between phases A and B 
passes the point $(J,J')$ = (1, 0.77) instead of (1, 4/7) in 
Fig. \ref{Fig_Phase}. 
      However the overall feature of Fig. \ref{Fig_Phase} well 
agrees with the numerical phase diagram. 

\begin{figure}[btp]
\begin{center}\leavevmode
\includegraphics[width=0.8\linewidth]{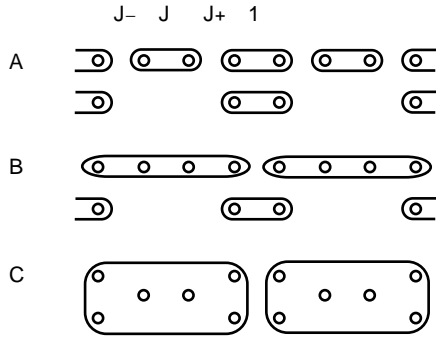}
\caption{ The SCS pictures for phases A, B and C. }
\label{Fig_SCS}
\end{center}
\end{figure}

      We explain the phases in Fig. \ref{Fig_Phase} in terms of 
singlet clusters (SC's). 
      An SC means a local singlet state formed by more than one 
$\frac{1}{2}$-spins. 
      We call a singlet state consisting of SC's 
the singlet cluster solid (SCS). 
      The SCS is an extension of the valence bond solid 
\cite{Affleck2} and special versions are used to explain phases 
for other spin chains \cite{Takano2,Chen}. 
      We show the SCS pictures for 
phases A, B and C in Fig. \ref{Fig_SCS}. 
      Here a spin with magnitude 1 is decomposed into two 
$\frac{1}{2}$-spins (circles) and the symmetrization 
retrieves the original spin. 
      A closed loop means an SC which is formed by 
$\frac{1}{2}$-spins in it. 
      In phase A, adjacent spins of magnitude 1 are strongly 
connected to form two valence bonds, since the coupling 1 
between them is larger than $J'$. 
      A pair of remnant $\frac{1}{2}$-spins form a valence bond 
almost irrespective of the value of $J$. 
      In phase C, both of $J$ and $J'$ are lager than 1 and hence four 
$\frac{1}{2}$-spins connected by them form an SC. 
      Phase B is an intermediate;  one of two valence bonds 
remains for a pair of spins of magnitude 1 and a 4-spin cluster 
is formed in a unit cell. 

      The SCS picture will be more generally explained \cite{Takano3}. 

\end{document}